\documentclass[%
 reprint,
superscriptaddress,
nofootinbib,
 amsmath,amssymb,
 aps,
]{revtex4-1}
\usepackage{tikz}
\usepackage{graphicx}
\usepackage{dcolumn}
\usepackage{bm}
\usepackage{comment}
\usepackage{notes2bib}
\bibnotesetup{
note-name = ,
use-sort-key = false
}

\begin{document}


\title{Wrapping and unwrapping multifractal fields}

\author{Samy Lakhal}
\email{samy.lakhal@outlook.fr}

 \affiliation{LadHyX, {UMR} CNRS 7646, Ecole Polytechnique, 91128 Palaiseau Cedex, France}
  \affiliation{Institut Jean Le Rond d’Alembert, UMR CNRS 7190, Sorbonne Universit\'e, 75005 Paris, France}
   \affiliation{Chair of EconophysiX \& Complex Systems,  Ecole Polytechnique, 91128 Palaiseau Cedex, France}

\author{Laurent Ponson}

\affiliation{Institut Jean Le Rond d’Alembert, UMR CNRS 7190, Sorbonne Universit\'e, 75005 Paris, France}

\author{Michael Benzaquen}
 \affiliation{LadHyX, {UMR} CNRS 7646, Ecole Polytechnique, 91128 Palaiseau Cedex, France}
  \affiliation{Chair of EconophysiX \& Complex Systems,  Ecole Polytechnique, 91128 Palaiseau Cedex, France}
 \affiliation{Capital Fund Management, 23 rue de l'Universit\'e, 75007 Paris, France}

\author{Jean-Philippe Bouchaud}
\affiliation{Chair of EconophysiX \& Complex Systems,  Ecole Polytechnique, 91128 Palaiseau Cedex, France}
 \affiliation{Capital Fund Management, 23 rue de l'Universit\'e, 75007 Paris, France}
\affiliation{Acad\'emie des Sciences, Quai de Conti, 75006 Paris, France}

\date{\today}

\begin{abstract}

We develop a powerful yet simple method that generates multifractal fields with fully controlled scaling properties. Adopting the Multifractal Random Walk (MRW) model of Bacry~{\it et al.}~\cite{bacry_multifractal_2001}, synthetic multifractal fields are obtained from the fractional integration of non-Gaussian fluctuations, built by a non-linear transformation of log-correlated Gaussian fields. The resulting fields are parameterized by their roughness exponent $H$, intermittency $\lambda$ and multifractal range $\xi_\omega$. We retrieve all the salient features of {the MRW, namely a quadratic scaling} exponent spectrum $\zeta_q$, fat-tail statistics of fluctuations, and spatial correlations of local volatility. {Such features can be finely tuned, allowing for the generation of ideal multifractals mimicking real multi-affine fields.} The construction procedure is then used the other way around to unwrap experimental data -- here the roughness map of a fractured metallic alloy. Our {analysis evidences subtle} differences with synthetic fields, namely anisotropic filamental clusters reminiscent of dissipation structures found in fluid turbulence.
\end{abstract}

\maketitle

Whether in the oceans~\cite{hersbach1999improvement}, in the sky~\cite{labini1998scale,coleman1992fractal,hentschel1984relative}, or in the mountains~\cite{mandelbrot1982fractal}, scale invariance is everywhere around us. This feature, formalized and popularized by Mandelbrot under the concept of fractals~\cite{mandelbrot2002gaussian,mandelbrot2013multifractals,feder2013fractals}, has reached a general audience, who can recognize and appreciate their unfathomable beauty. For physicists, systems are said {to be} fractal when their observables display deterministic or statistical invariance under affine transformations of their space-time variables. For a fluctuating or stochastic field $h(\bm{r}),\ \bm{r} \in \mathbb{R}^d$, this property corresponds to the scale invariance of the moments of its increments, which translates as the following variogram property:
\begin{equation}
V_q(\delta {r}) = \langle |h( \bm{r} + \bm{\delta r}) - h( \bm{r}) |^q\rangle_{\delta r = \|\bm{\delta r}\|} 
\sim \delta r^{\zeta_q},
    \label{eq:scaling}
\end{equation}
where $\langle .\rangle$ denotes the spatial or ensemble average, and $\zeta_q$ is a \textit{scaling exponent spectrum}. A linear spectrum $\zeta_q = qH$ with Hurst exponent $H$ {corresponds to a} monofractal scaling, usually found in linear systems.
Instead, multifractals are characterized by a non-linear spectrum $\zeta_q\nsim qH$, reported in forced non-linear systems~\cite{chibbaro_elastic_2016,boudaoud2008observation}, and which may originate from a wide range of mechanisms such as self-organization~\cite{bak1987self}, exotic fluctuations~\cite{barabasi_multifractality_1991}, or hierarchical interactions~\cite{foias1988modelling}. 
For the latter, a classic example may be of fluid turbulence~\cite{K41,frisch_turbulence_1995,mandelbrot1999intermittent}, in which scale invariance originates from the cascading of energy from large eddies at the injection/forcing scale down to mesoscopic dissipation scale. 

Looking back, our ability to generate random monofractal objects~\cite{fournier1982computer,pesquet2002stochastic,moreva2018fast} has allowed scientists to better understand their emergence in natural phenomena~\cite{mandelbrot1982fractal}, and led to many applications, for which a non-exhaustive list may include roughness characterization methods~\cite{schmittbuhl1995reliability,morel2023scale}, simulations of frictions~\cite{yastrebov2015infinitesimal}, deformations~\cite{yan1998contact} or flows~\cite{talon2010permeability,liu2015fractal} on fractal landscapes, and even in the training of learning-based image recognition algorithms~\cite{hirokatsu2022pre,anderson2022improving,li2022robust}. The wide diversity of phenomena displaying scaling invariance makes of fractal generation algorithms an essential tool for physicists, but also for graphics engineers~\cite{barnsley1988fractals,tessendorf2001simulating} and artists, see e.g. the self-similarity structures of Pollock's paintings or of traditional middle-eastern ornaments~\cite{bridges2003:1,taylor2002order}.
Yet, while synthetic multifractals too could benefit from various applications, generation methods are scarce and generally limited to one dimensional data (see~\cite{morel2023scale} and refs therein). Indeed, sampling methods for multifractal random fields in $d>1$ suffer from several limitations, related to the symmetry~\cite{LOVEJOY20101393,gagnon2003multifractal}, isotropy and continuity of their statistics~\cite{decoster2000wavelet}.

\begin{figure}[b]
    \centering
    \includegraphics[width = 1.\linewidth]{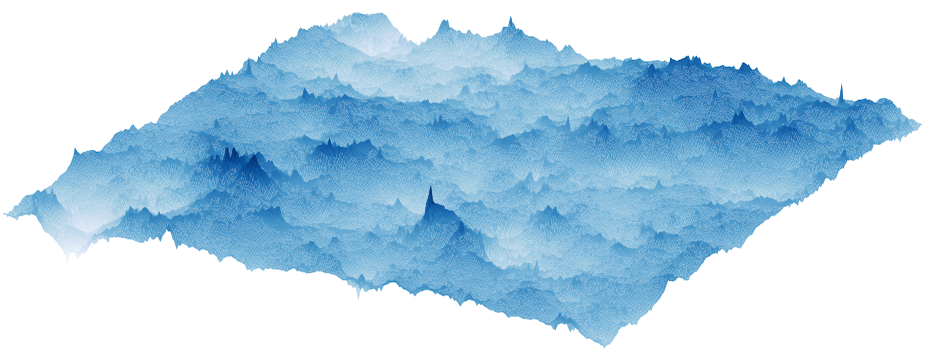}
    \caption{Synthetic multifractal field in dimension $d=2$.}
    \label{fig:MRW3D}
\end{figure}

In this article, we provide a powerful, yet simple method for the generation of multifractal random fields in any dimensions (see Fig.~\ref{fig:MRW3D} for an example in $d=2$). Our model builds on the work of Bacry {\it et al.}~\cite{bacry_multifractal_2001} who defined a multifractal process with continuous dilation invariance properties, and that we generalize to higher dimensions by using fractional operators.
In practice, as described in the first part of our article, our method consists in passing a log-correlated Gaussian field $\omega$ through three distinct steps of (i) exponentiation, (ii) symmetrization and (iii) fractional integration, resulting in a multifractal field $h(\bm{r})$ with fully controlled scaling properties. This construction allows for the prediction of the density probability of increments at any scale $\ell < \xi_\omega$, from the parameters $\lambda$ and $H$ alone.~The second part of the article is devoted to the analysis of an experimental multifractal field, here the height map $h(\bm{r})$ of a fractured metallic alloy. By applying our construction steps backwards, we \textit{unwrap} $h(\bm{r})$ and extract the local volatility $\hat \omega$, from which multiscaling originates. We evidence subtle differences between synthetic $\omega$ and experimental $\hat\omega$ fields, providing insights on the mechanism terminating the cascading processes in fracture problems. We conclude our article by discussing the implications of our method for the {investigation} of strongly-coupled processes leading to multifractality.

\textit{Monofractal fields --} Monofractal Gaussian fields can be defined from the application of fractional operators to white Gaussian noise. In dimension $d=1$, these operators result from classic integration and derivation~\cite{mandelbrot_fractional_1968}. For $d>1$, one {may enforce isotropy and translation invariance by introducing the}  Fractional Laplacian $(-\Delta)^\alpha$~\cite{lischke_what_2020} and define a monofractal field  $\omega(\bm r)$ ($\bm r \in \mathbb{R}^d$) from:
\begin{equation}
    \omega(\bm r) = (-\Delta)^{-\frac{H+d/2}{2}}\eta(\bm r),
    \label{eq:FractionalIntegration}
\end{equation}
where $\eta$ is a white Gaussian noise of dimensions $d$. In Fourier space, this operation amounts to using the filter $G(\bm{k}) = 1/k^{H+d/2}$ ($k= \|\bm{k}\|$), which possesses scale and rotation invariance\bibnote{In a lattice space of step $1$, the Laplacian is expressed in terms of nearest neighbours increments, which is equivalent in the Fourier space to replacing $\|\bm{k}\|^2$ by $2\times \sum_{i=1}^d (1-\cos k_i)$.}. We may distinguish two main families of power-law correlated fields. For $H \in [-d/2,0[$, $\omega(\bm r)$ is a zero mean stationary Gaussian field with decaying power-law correlations $C_{\omega}(\bm{\delta r} ) =\langle \omega(\bm{r} +\bm{\delta r})\omega(\bm{r}) \rangle \propto \delta r^{2H}$. For $H \in ]0,1]$, $\omega(\bm r)$ is a fractional Gaussian field~\cite{bojdecki_fractional_1999,lodhia_fractional_2016} of monofractal scaling $V_q(\bm{\delta r})\propto {\delta r}^{qH}$. 
The particular case $H=0$, sitting at the transition point between the two cases mentioned above, corresponds to system size {and regularization}-dependent logarithmic scaling.  

In order to mimic natural fractals that only expand over a finite range of length scales, we introduce an upper cutoff $\xi_\omega$ using the modified operator $(-\Delta)\to ({\xi_\omega^{-2}}-\Delta)$ that dampens long-range correlations. In Fourier space, this translates as $k^2 \to \left(2\pi/\xi_\omega\right)^2+k^2$, {which prevents the power-law scaling of the low frequency regime}. Simple calculations~\cite{gradshteyn2014table} provide the correlations of $\omega$, $C_{\omega}(\bm{\delta r})\propto \delta r^H K_{H}(\delta r/\xi_\omega)$, where $K_H$ is the modified Bessel function of the second kind with parameter $H$ We retrieve the classic Whittle-Mat\'ern correlations~\cite{whittle1954stationary} which behave similarly to unregularized fields at short scales $\delta r\ll \xi_\omega$, and decay exponentially such that $C_{\omega}(\bm{\delta r}) \propto e^{-\delta r/\xi_\omega}$ for $\delta r\gg \xi_\omega$. Taking $H = 0$ recovers logarithmic scaling $C_\omega(\bm{\delta r}) \propto -\log(\delta r/\xi_\omega) +\mathrm{k}$ over a finite range of length scales $\delta r < \xi_\omega$, and where $\mathrm{k} = \gamma - \log 2 \approx .07$ is a small correction term to the logarithmic asymptote of $K_0(u)$. We observe the effect of $\xi$ on the visual and statistical properties of $H=0$ fields in~\bibnote[Appendix]{See Appendix. below for a study of $H=0$ fields, a derivation of $\zeta_q$, an expression of $G_{\ell/L}(u)$, and a parameter estimation of sampled fields, which includes Refs.~\cite{duplantier_log-correlated_2017,hager_multiplicative_2022,neuman_fractional_2018,Riesz,bacry_multifractal_2001,benzi1993extended,chibbaro_elastic_2016,boudaoud2008observation}.}.

\begin{figure}
    \centering
    \includegraphics[width = .95\linewidth]{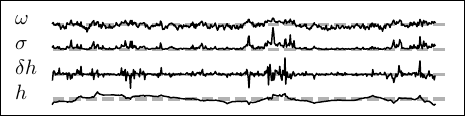}
    \caption{Construction of multifractal random signals, illustrated in dimension $d=1$. From top to bottom, (i) a log-correlated ($H=0$) field $\omega$ is exponentiated into $\sigma=e^{\omega}$, (ii) symmetrized into $\delta h = s\sigma$ and (iii) fractionally integrated into $h$. Horizontal lines indicate the origins $y=0$ of each curve.}
    \label{fig:Trajectories}
\end{figure}

\textcolor{black}{\textit{Synthetic multifractal fields -- }} We now describe the construction of multifractal fields, consisting of the three steps shown in Fig.~\ref{fig:Trajectories}.
First, one builds non-Gaussian fluctuations by taking the exponential of a Gaussian field $\omega$. The resulting field $\sigma = e^\omega$ displays log-normal statistics, and amplifies the largest fluctuations of $\omega$ while tempering the lowest ones, see Fig~\ref{fig:Trajectories}. Such signals, often referred to as crackling noise, are observed in elastic models driven in disordered media, where the largest fluctuations organize in bursts called avalanches~\cite{rosso2009avalanche,le2021spatial}. The  correlations of $\sigma$ can be directly computed from the characteristic functions of multivariate Gaussian distributions. In particular, taking $H=0$ with $C_\omega(\bm{\delta r})\approx-\lambda \log \delta r /\xi_\omega$ for $\delta r<\xi_\omega$, one gets
\begin{equation}
     \left\langle \sigma(\bm{r_1})\dots\sigma(\bm{r_{q}})\right\rangle =\prod_{1 \leq i \leq j \leq q} \|\bm{r_i} - \bm{r_j}\|^{-\lambda},
    \label{eq:LogNormalScaling}
\end{equation}
where $\bm{r_1},\dots, \bm{r_q} \in \mathbb{R}^d$, and $\|\bm{r_i} - \bm{r_j}\|<\xi$ for all $(i,j)$.
Here, the \textit{intermittency coefficient} $\lambda$ and integral range $\xi_\omega$ control the strength and the spatial extension of the bursts of $\sigma$ respectively. The resulting process is a \textit{log-normal continuous cascade}~\cite{BacryContinuous2013}, and is scale invariant since applying $\bm{r_i} \to \gamma \bm{r_i}$ multiplies Eq.\eqref{eq:LogNormalScaling} by $\gamma^{-\frac{\lambda}{2} q(q-1)}$. We will see that this quadratic scaling in $q$ directly influences the shape of $\zeta_q$. However, we note that the statistics of $\sigma$ are skewed, and do not reproduce the symmetry observed in experimental data, e.g. turbulence velocity fields~\cite{chevillard2005intermittency,chevillard2006unified}. The second step, introduced by Bacry, Delour and Muzy for the Multifractal Random Walk (MRW)~\cite{bacry_multifractal_2001} consists in symmetrizing the non-Gaussian fluctuations by multiplying them by a zero-mean white Gaussian noise $s$. The resulting field $\delta h = s. \sigma$ possesses symmetrical statistics, where $\sigma$ defines an intermittent volatility envelope in which $\delta h$ fluctuates, see Fig.~\ref{fig:Trajectories}. The third and last step consists in integrating $\delta h$. For $d=1$, $\delta h$ exactly defines the increments of the MRW and an integration retrieves a multifractal signal. For $d>1$, we sample multifractal fields by using fractionally integrating $\delta h$, into
$h:= (-\Delta)^{-\frac{H+d/2}{2}}\delta h$, using Eq.~\ref{eq:FractionalIntegration}. Similarly to the construction proposed in Ref.~\cite{duchon2005champs}, these fields display an asymptotic ($\delta r\ll \xi_\omega)$ multifractal scaling of exponent spectrum:
\begin{equation}
    \zeta_q = qH - \frac{\lambda}{2}q(q-2),
    \label{eq:MRWspectrum}
\end{equation}
a result that we derive in~\bibnotemark[Appendix]. In $d=1$, taking $H=1/2$ provides the MRW. In the general case, the constructed field recovers monofractality with  $\zeta_q = qH$ for $\delta r > \xi_\omega$. Note that introducing a second cut-off $\xi_h$ in the last integration step leads to a saturation of the variograms $V_q$ for $\delta r > \xi_h$, as observed in numerical and experimental data~\cite{chibbaro_elastic_2016,boudaoud2008observation}. 

\begin{figure}[b]
    \centering
    \includegraphics[width = .9\linewidth]{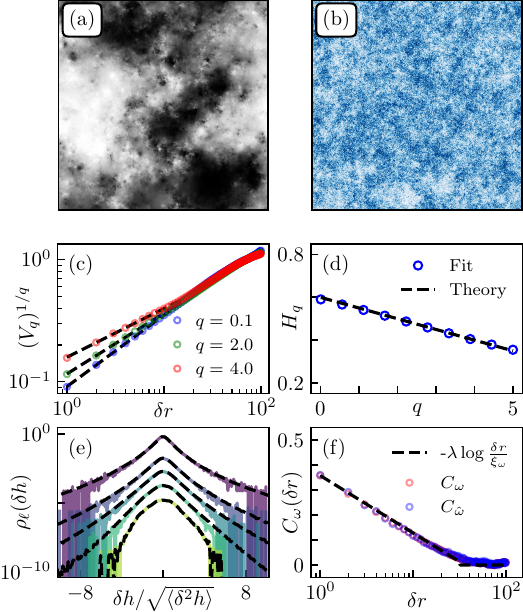}
    \caption{Analysis of a synthetic multifractal field, with dimension $d=2$, size $512\times 512$, and parameters $(H,\lambda,\xi_\omega) = (0.5,0.1,32)$. (a) $h(\bm r )$. (b) $\omega(\bm r )$.  (c) $(V_q)^{1/q}$ vs $\delta r$, with power law fits. (d) $H_q = \zeta_q /q$ vs $q$, fitted from variograms, with theory from Eq.~\eqref{eq:MRWspectrum}. (e) p.d.fs $\rho_\ell$ vs $\delta h$, for $\ell/\xi_\omega = 1/32, 1/16, 1/8, 1/4, 1/2 \text{ and } 1$ (top to bottom, shifted for visibility). Theoretical predictions from Eq.~\eqref{eq:Castaing} in black, with reference scale $L = \xi_\omega$. (f) Spatial correlations of $\omega$ and $\hat \omega$, with theoretical logarithmic asymptote.}
    \label{fig:MultifractalAnalysis}
\end{figure}

\textit{Intermittency characterization --} We verify the {effectiveness} of our method by generating surfaces of size $512\times512$ pixels, as the one shown in Fig.~\ref{fig:MultifractalAnalysis}(a) for $H=0.5$, $\lambda=0.1$ and $\xi_\omega = 32$.  
We first check its multifractal properties: in Fig.~\ref{fig:MultifractalAnalysis}(c), we analyze the multifractal scaling of $h$ by computing the power-law exponents of the rescaled variograms  $(V_q)^{1/q}$ in the multifractal regime $\Delta r<\xi_\omega$. We observe in Fig.~\ref{fig:MultifractalAnalysis}(d) that the generalized Hurst exponents $H_q = \zeta_q/q$ follows the linear behavior $H_q =\zeta_q/q =  H - \lambda/2\, (q-2)$ expected from Eq.~\eqref{eq:MRWspectrum}. Note that the multi-affine to mono-affine crossover $\xi_\omega$ is evidenced from the collapse of the rescaled variograms $(V_q)^{1/q}$ for $r > \xi_\omega$.  

As shown in the following and as previously reported in Refs.~\cite{zamparo_apparent_2017,bouchaud_apparent_1999}, the analysis of $h$ only may be insufficient to ascertain multifractal properties. This difficulty can be circumvented by studying directly the field $\omega$ introduced previously. For MRW, $\lambda$ and $\xi_\omega$ can be measured from the local log-volatility field $\hat \omega_\epsilon = \log|\delta_\epsilon h|$, {also called \textit{magnitude}~~\cite{bacry_modelling_2001,muzy_intermittency_2010,arneodo1998analysis} or log-dissipation rate in turbulence}. For multifractal fields, $h$ can be unwrapped using the operator~\cite{LOVEJOY20101393,gagnon2003multifractal}:
\begin{equation}
    \hat \omega(\bm r) = \log |(-\Delta)^{\frac{H+d/2}{2}}h(\bm r)|,
    \label{eq:Omega}
\end{equation}
{which differs from previous studies~\cite{vernede_turbulent_2015,o1993spatial} by fully unwrapping the effect of the roughness exponent $H$ in the definition of the magnitude.}
The obtained log-volatility field, shown in Fig.~\ref{fig:MultifractalAnalysis}(b), displays long-range correlations reminiscent of the multifractal properties of the original field $h(\bm{r})$ shown in Fig.~\ref{fig:MultifractalAnalysis}(a). The fit of the correlations of $\hat \omega(\bm r)$ with a logarithmic slope retrieves
provides  $\lambda = 0.15$ and $\xi_\omega = 32$, perfectly matching their prescribed value. In order to assess the versatility of our method, we generate a wide variety of multifractal fields of size $512 \times 512$ with prescribed properties in the range $0 \leq \lambda \leq 0.5$ and $0 \leq H \leq 1$ for a fixed cut-off length $\xi_\omega = 30$ in~\bibnotemark[Appendix]. The parameter values $\lambda$ and $H$ are then measured either from the volatility field $\hat \omega(\bm r)$ or the height field $h(\bm r)$. We find a good agreement with the prescribed values, especially when the $\hat{\omega}$-field is used, suggesting that the study of the spatial correlations of the volatility field is a more direct and accurate way to characterize multifractal fields.

The multifractal scaling directly implies that increments $\delta_{\bm \ell } h =h(\bm r+\bm \ell ) - h(\bm r) $ are linked together through a cascading rule, defined as the ratio of fluctuations $W_{\ell/L}=\delta_\ell h/\delta_L h $, and such that $\langle W_{\ell/L}^q\rangle = (\ell/L)^{\zeta_q}$. Here, the cascading rule depends on the scale ratios exclusively, hence describing a scale-invariant cascade. 
As proposed by Castaing~{\it et al.}~\cite{castaing_velocity_1990}, this relation can be used to link the probability density functions (p.d.f) $\rho_{\ell}(\delta h)= \mathbb{P}(\delta_\ell h=\delta h)$ through the following dilation invariance:
\begin{equation}
\rho_{\ell}(\delta h)=\int G_{\ell/L}(u) e^{-u} \rho_{L}(e^{-u}\delta h) d u.
    \label{eq:Castaing}
\end{equation}
The kernel $G_{\ell/L}(u)$ is the Gaussian p.d.f of $\log W_{\ell/L}$, that depends on $\zeta_q$, and thus on $H$ and $\lambda$ only. Its analytical expression is provided in~\bibnotemark[Appendix]. In practice, the statistics of fluctuations is computed at the scale $L = \xi_\omega$ from which the statistics at smaller scales is inferred using Eq.~\eqref{eq:Castaing}. This construction provides an alternative characterization of the multifractal behaviour of $h$, as it captures quantitatively the ever stronger departure from Gaussianity as we go deeper into the multifractal regime. 
In Fig.~\ref{fig:MultifractalAnalysis}(e), we compute the distributions $\rho_\ell(\delta h)$ on our synthetic surfaces and observe the expected transition from fat to Gaussian tail as $\ell$ increases. The numerical data is in perfect agreement with Eq.~\eqref{eq:Castaing}.

\begin{figure}[b]
    \includegraphics[width = .9\linewidth]{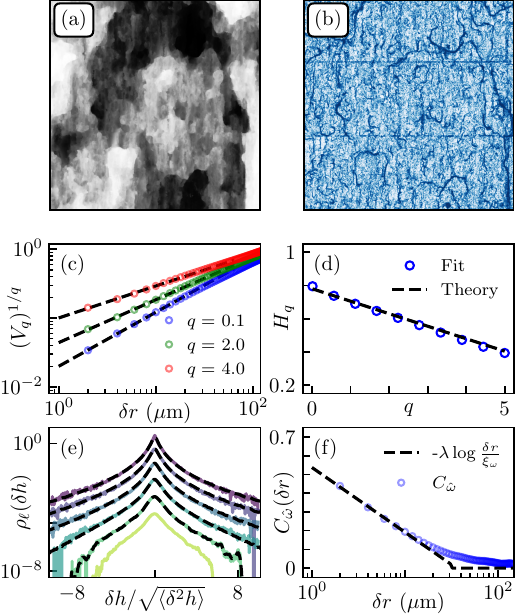}
    \caption{Unwrapping of an experimental multi-affine field. (a) Height map $h(\bm r)$ of a fractured metallic alloy of size $2\times2~\mathrm{mm}^2$ with $2~\mu \mathrm{m}$/pixel. (b) $\hat \omega(\bm r)$, computed from Eq.\eqref{eq:Omega}. (c) $(V_q)^{1/q}$ vs $\delta r$, with power-law fits. (d) $H_q = \zeta_q /q$, fitted from variograms, with theory from Eq.~\eqref{eq:MRWspectrum} using $H=0.63$ and $\lambda=0.15$. (e) p.d.fs $\rho_\ell$ vs $\delta h$, with $\ell/\xi_\omega = 1/16, 1/8, 1/4, 1/2, 1 \text{ and } 2$. Theoretical predictions from Eq.~\eqref{eq:Castaing} in black, with reference scale $L = \xi_\omega$. (f) Correlations of $\hat \omega$, with fitted logarithmic asymptote, retrieving $\lambda = 0.15$ and $\xi_\omega = 33\mu\mathrm{m}$.}
    \label{fig:FractureAnalysis}
\end{figure}

\textit{Application to experimental data --} We now consider experimental multifractal data, here the height map of the surface of a fractured metallic alloy measured by interferometric profilometry (see Fig.~\ref{fig:FractureAnalysis}(a)). Fracture surfaces are archetypes of multi-affine fields~\cite{vernede_turbulent_2015}, even though the physical origin of their complex geometry is still debated~\cite{santucci2007statistics,ponson2021unified,bouchbinder2006fracture}.
The analysis carried before on synthetic fields is implemented in Fig.~\ref{fig:FractureAnalysis} to the experimental fracture surface, leading to the parameter values $(H,\lambda,\xi_\omega,\xi_h) = (0.63,0.15,33\mu\text{m},360\mu\text{m})$. As reported in \cite{vernede_turbulent_2015}, we recover all the salient features of our MRW-based multifractal fields, namely a linear decay $H_q = H - \lambda/2 \, (q-2)$ of the exponents and a log-correlated $\hat{\omega}$-field. Notice however in Fig.~\ref{fig:FractureAnalysis}(c) the slow transition towards monofractal scaling, a feature that also manifests in the statistics of height fluctuations shown in Fig.~\ref{fig:FractureAnalysis}(e) that show significant deviations to Gaussian statistics even for $\ell \simeq 2 \xi_\omega$. Such soft crossover towards a Gaussian mono-affine behavior may result from the particularly marked cliff-like patterns of the fracture surface, a feature that is investigated in more details below.

\begin{figure}[h]
    \centering
    \includegraphics[width = .9\linewidth]{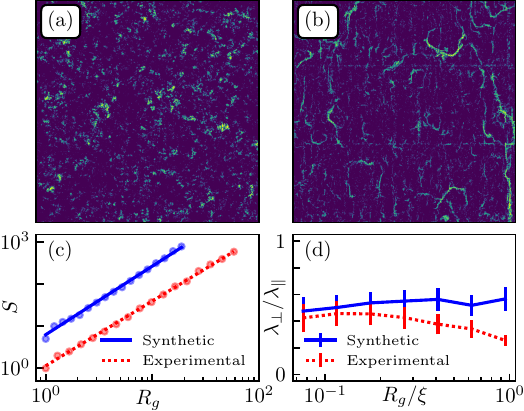}
    \caption{Cluster analysis of $\omega$ fields. (a) and (b) Volatility fields of artificial and experimental surfaces, thresholded at $10$\% of their largest values. (c) $S$ vs $R_g$, fitted by power-law exponents $D = 1.65 \pm.03$ and $1.53\pm.02$ for synthetic and experimental clusters respectively. (d) $\lambda_\parallel/\lambda_\perp$ vs $R_g/\xi$, with
    error bars corresponding to standard deviations.}
    \label{fig:Clusters}
\end{figure}

The cliff-like organization of the experimental surface is highlighted in Fig.~\ref{fig:Clusters}(b) that shows the $10\%$ largest values of $\hat \omega$. The most intermittent clusters organize in filamental structures; these are visually more compact for the synthetic field shown in  Fig.~\ref{fig:Clusters}(a). For fracture surfaces, this property originates from the existence of local damage mechanisms, culminating through the formation of mesoscale structures of size $\xi_\omega$. To explore these differences, we first compute the fractal dimension $D$ of these clusters~\cite{vernede_turbulent_2015}, as shown in Fig.~\ref{fig:Clusters}(c). Their area $S$ and their spatial extension $R_\mathrm{g}$ defined as the gyration radius scales as $S = R_\mathrm{g}^D$ for both types of clusters with nearly the same exponent, even though the fractal dimension for the synthetic surface (that display more compact features) is slightly larger ($D \approx 1.65$ instead of $1.53$).  We then go one step further in~Fig.~\ref{fig:Clusters}(d) by comparing the clusters' aspect ratio, defined as the ratio $\lambda_\perp/\lambda_\parallel<1$ of the two eigenvalues of their inertia tensor. The lower aspect ratio of the experimental clusters reveals their filamental structure, specially for large $R_g$. Such topology is reminiscent of kinetic energy dissipation bursts in fluid turbulence and is related to cliff-like regions on the fracture surface. Understanding the emergence of such regions from the cooperative dynamics of damage coalescence mechanisms taking place within the \textit{fracture process zone} may shed light on the microscopic origin of the fracture energy of materials~\cite{mayya2022criticality}.

\textit{Conclusion and discussion --} We now summarize our main findings and discuss their implications for the study of multifractal phenomena. We first generated monofractal Gaussian fields of Hurst exponent $H$ and fractal range $\xi_\omega$ from which we built log-normal cascades by exponentiation in the case $H=0$. This constitutive brick was then used to compute symmetric non-Gaussian fluctuations $\delta h$ whose fractional integration led to synthetic multifractal random fields with quadratic scaling exponents~$\zeta_q$. The fields generated by our method retrieve all the salient features of {classic multifractal random walks}: {quadratic scaling exponent spectrum}, log-correlated volatility and a transition from fat to Gaussian tail statistics. Our method is limited here to the generation of isotropic multifractals, but the anisotropy  observed in many experimental systems, in particular fracture surfaces~\cite{ponson2006two,bouchbinder2005disentangling}, can be retrieved using anisotropic kernels. Also, our method provides multiscaling asymptotically, but exact multiscaling may be recovered by domain warping~\cite{lagae2010survey} of fractional Brownian fields with multifractal measures~\bibnote{Circulant embedding and real-time resampling~\cite{vanlawick1996} also constitute possible alternatives to generate multifractals with minor discretization and finite-size effects.}.

We believe an important contribution is the idea of applying our methodology ``backwards'' such as to {\it unwrap} experimental multifractal fields and identify the singularities responsible for multifractality. In the case of the height map of a fractured material, we identify and characterize all the basic ingredients used to construct synthetic fields, but also highlight some fundamental differences, such as the softer crossover towards monofractality (see Fig.~\ref{fig:FractureAnalysis}(f)) and the non-trivial topology of the most intermittent bursts (see Fig.~\ref{fig:Clusters}(d)). This last property is reminiscent of the cascading mechanism observed in turbulence, and during which vorticity filaments near dissipation scales~\cite{kraichnan1980two,boffetta2012two}. It suggests that during material failure, cooperative coalescence of damage cavities take place and culminates in the formation of large-scale cliff-like filament structures. {A next step in that investigation may rely on the investigation of fuse-based~\cite{alava2006statistical} or coagulation-based descriptions~\cite{ferreira2021stationary}, where such cavities are continuously absorbed and created in the vicinity of the crack tip.} Beyond material failure, it implies that a rich and insightful information is embedded in the topology of these highly intermittent clusters, beyond the standard scaling exponents of (multi)fractals.

\begin{acknowledgments}  
We thank Rudy Morel for very fruitful discussions. S.L thanks Guillaume De Luca for the visualization algorithm of Fig.~\ref{fig:MRW3D}.
This research was conducted within the Econophysics \& Complex Systems Research Chair, under the aegis of the Fondation du Risque, the Fondation de l'Ecole polytechnique, the Ecole polytechnique and Capital Fund Management. 
\end{acknowledgments}


\bibliographystyle{apsrev4-1}
\bibliography{Lakhal2023Wrapping}

\begin{thebibliography}{76}
\bibitem [{1()}]{1}%
  \BibitemOpen
  \href@noop {} {}\bibinfo {note} {In a lattice space of step $1$, the Laplacian is expressed in terms of nearest neighbours increments, which is equivalent in the Fourier space to replacing $\|\bm {k}\|^2$ by $2\times \sum _{i=1}^d (1-\cos k_i)$.}\BibitemShut {Stop}%
\end{thebibliography}

\pagebreak
\appendix
\section{Gaussian random fields, the case H=0 \label{app:H=0}}
 We illustrate the effect of $\xi_\omega$ on the properties of sampled Gaussian random fields in $d=2$, and for $H=0$~\footnote{In dimension $d=2$, the $H=0$ case corresponds to both log-correlated and free Gaussian field with Lagrangian $\mathcal{L}[\phi]= (\nabla \phi)^2 +\phi^2/\xi_\omega^2$~\cite{duplantier_log-correlated_2017,hager_multiplicative_2022,neuman_fractional_2018}.}.

\begin{figure}[h]
    \centering
    \includegraphics[width = 1.\linewidth]{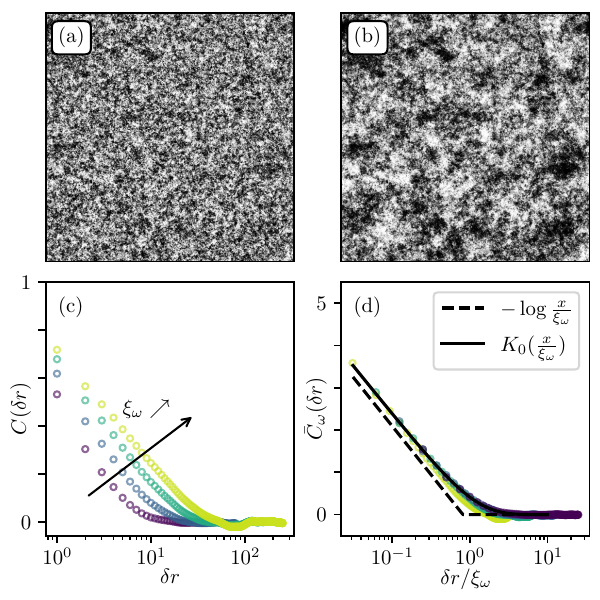}
    \caption{Regularization influence of $H=0$ Gaussian fields for $d=2$. (a), (b) Quantile representation of $512\times 512$ samples for $\xi_\omega \in \{4,32\}$ respectively. (c) Isotropic autocorrelation functions for $\xi_{\omega}=4,8,16,32$. (d) Correlations rescaled by $\xi_\omega$, showing good collapse around the theoretical correlations $C_\omega(\delta r) \propto K_0(x/\xi))$ in black. The logarithmic asymptote in black dotted lines is shifted for visibility.}
    \label{fig:LogCorrelations}
\end{figure}

In Fig.~\ref{fig:LogCorrelations}(a) and (b), $\xi_\omega$ directly influences the range of correlations, and consequently the visuals. The correlation functions in (c), possess $\xi$-dependent logarithmic scaling, and can be conveniently collapsed in (b). The logarithmic asymptote and the theoretical correlation $K_0(u)$ are shown in the same figure. This ability to control the logarithmic range will be useful to tune the multifractal properties of sampled fields, seen later.

\section{Scaling of synthetic multifractal fields: stationarity and asymptotic scaling.\label{appendix:ScalingProof}}
In this appendix, we consider a field $h$ obtained from the construction described in the main text. We demonstrate that the jumps, $\Delta_{\bm{\ell}} h(\bm r) = h(\bm r +\bm{\ell}) - h(\bm r)$ describe a stationary process whose moments scale with $\ell$. We recall that $h(\bm r), \bm r\in \mathbb R^d$ is built from the fractional integration of $\delta h(\bm r) = s_(\bm r)e^{\omega(\bm r)}$. We consider Gaussian statistics for $s$ and $\omega$ and the following correlations:
\begin{equation}
  \left\{
      \begin{aligned}
        C_s(\bm r) &\propto \delta^{d}(\bm r)\\
        C_\omega (\bm r)&\approx -\lambda \log\frac{r}{\xi_\omega},\ r< \xi_\omega.
      \end{aligned}
    \right.
\end{equation}
The relation $h(\bm r)=(-\Delta)^{-\frac{H+d/2}{2}}\delta h(\bm r)$ rewrites in terms of Riesz potential~\cite{Riesz} as:
\begin{equation}
    h(\bm r)=\int_{\mathbb{R}^d}d\bm{r_1}\frac{\delta h(\bm r+\bm{r_1})}{\|\bm{r_1}\|^{\frac{d}{2}-H}}.
\end{equation}
From now on, integrals will always be over the whole domain $\mathbb{R}^d$, unless specified

\subsection{Stationarity of increments}
We calculate the q-th order moments of jumps between $\bm{r}-\frac{\bm\ell}{2}$ and $\bm r +\frac{\bm\ell}{2}$, with $\bm\ell = \ell\bm{e}$, ($\|\bm{e} \|=1$) and $q=2m$. Using the symmetry of the kernel, the order $q$ variogram writes as:
\begin{multline}
    \langle |h(\bm{r}+\frac{\bm\ell}{2})-h(\bm{r}-\frac{\bm\ell}{2})|^{q}\rangle =
     \int d\bm{r_1}\dots\int d\bm{r_q}\\
     \prod_{i=1}^q \left[ \|\bm{r}+\frac{\bm\ell}{2} -\bm{\bm{r_i}}\|^{H-\frac{d}{2}} - \|\bm{r}-\frac{\bm\ell}{2} -\bm{r_i}\|^{H-\frac{d}{2}}\right] \\
     \times\langle\prod_i^q \delta h(\bm{r_i})\rangle.
     \label{eq:appJumps}
\end{multline} 
The fields $s$ and $\omega$ being independent, one gets:
\begin{equation}
\langle\prod_i^q \delta h(\bm{r_i})\rangle = \langle \prod_{i=1}^q s(\bm r_i) \rangle \langle e^{\sum_i^q\omega(\bm r_i)} \rangle,
\end{equation} 
which becomes, using the Wick theorem and the characteristic function of multivariate Gaussian processes leads to:
\begin{multline}
    \langle\prod_i^q \delta h(\bm{r_i})\rangle = \prod_{i_1,\dots,i_{2m}}C_s(\bm{r_{i_1}}-\bm{r_{i_2}} )    \times  \\
    \dots \times C_s(\bm{r_{i_{2m-1}}}-\bm{r_{i_{2m}}})\\
    \times  B e^{\frac{1}{2}\sum_{i\neq j=1}^{2m}\frac12 C_\omega(\bm{r_{i}}-\bm{r_{j}} )},
\end{multline}
where $B$ is the variance/diagonal term of the correlation matrix, unimportant here as we observe the scaling exclusively. 
Injecting this last expression in the first integral and applying the change of variable $\bm{r_i}\to\bm{r_i} +\bm{r}$ suppresses $\bm{r}$ from the expression. The functional now exclusively depends on $\bm \ell$, which implies stationarity, or translation invariance. This translates as the following:
\begin{align}
    V_q(\bm{\ell}) &=\langle |h(\bm{r}+\frac{\bm\ell}{2})-h(\bm{r}-\frac{\bm\ell}{2})|^{q}\rangle\\
    &=\langle |h(\bm{\ell})-h(\bm{0})|^{q}\rangle.
\end{align} 
We also note the rotation invariance of $V_q(\bm \ell )$.

\subsection{Derivation of the scaling exponent spectrum $\zeta_q$} We derive the scaling exponent spectrum of jumps $h(\bm{\ell})-h(\bm{0})$ in the $\ell \ll \xi_\omega$ limit of Eq.~\eqref{eq:appJumps}. The moments of order $q=2m$ write:
\begin{multline}
    V_q(\bm{\ell}) =
     \int d\bm{r_1}\dots\int d\bm{r_q}\\
    \prod_{i=1}^q \left[ \|\bm{r}_i-\frac{\bm\ell}{2} \|^{H-\frac{d}{2}} - \|\bm{r}_i+\frac{\bm\ell}{2}\|^{H-\frac{d}{2}}\right] \\
    \times \langle\prod_i^q \delta h(\bm{r_i})\rangle
     \label{eq:appJumps2}
\end{multline} 

Using that $s$ is $\delta$-correlated, we identify successive terms $\bm{r_{2i-1}}=\bm{r_{2i}}$ and get:
\begin{multline}
    V_q(\bm \ell ) =
     \int d\bm{r_2}\dots\int d\bm{r_{2m}}\\
     \prod_{i=1}^{m}\left[ \|\bm{r_{2i}}-\frac{\bm\ell}{2} \|^{H-\frac{d}{2}} - \|\bm{r_{2i}}+\frac{\bm\ell}{2}\|^{H-\frac{d}{2}}\right]^2\\ \times B \prod_{i\neq j = 1}^m e^{-{\frac{1}{2}C_{\omega}(\bm{r_{2i}}-\bm{r_{2j}})}}.
\end{multline} 

Note that this integral is well-defined for $H<1$. Similarly to \cite{bacry_multifractal_2001}, we make the change of variable $\bm{r}_i = \ell \bm{u}_i$ and retrieve: 


\begin{multline}
    V_q(\bm \ell ) \propto
     \ell^{2mH}\int d\bm{u_2}\dots\int d\bm{u_{2m}}
     \\
     \prod_{i=1}^{m}\left[ \|\bm{u_{2i}}-\frac{\bm e}{2} \|^{H-\frac{d}{2}} - \|\bm{u_{2i}}+\frac{\bm e}{2}\|^{H-\frac{d}{2}}\right]^2\\ \times \prod_{i\neq j = 1}^m e^{-\frac12 {C_{\omega}(\ell(\bm{u_{2i}}-\bm{u_{2j}}))}}.
\end{multline}   

To recover the scaling of this last integral term, we split the integration domain into the reunions of subsets $\| \bm{u_i} - \bm{u_j}\|<\xi_\omega/\ell$. This allows one to extract the logarithmic correlations of $C_\omega$, leading to a dominant term of the form $\ell^{-2\lambda m (m-1)}$. 

Power counting all contributions finally leads to $V_{2m}(\bm \ell )\propto \ell^{2mH -2\lambda m(m-1)}$, which recovers:
\begin{equation}
    V_{q}(\bm \ell ) \underset{\ell \ll \xi_\omega}{=} K_{q}\ell^{\zeta_{q}}, 
\end{equation}
where $K_{q}$ depends on the standard deviation of $s$ and $\zeta_q$ is the scaling exponent spectrum of expression:
\begin{equation}
    \zeta_q = qH -\frac{\lambda}{2}q(q-2),
\end{equation}
which corresponds to the MRW scaling for $H=1/2$. Finally, note that this scaling can be generalized to all $q$ from analytical continuation arguments.\\

\section{Self-similarity kernel $G_{\ell/L}(u)$ and fluctuation ratio $W_{\ell/L}$ distribution\label{app:SelfSimilarityKernel}}

The fluctuation ratio $\langle W_{\ell/L}^q\rangle = (\ell/L)^{\zeta_q}$ is characterized by its moments:
\begin{equation}
    \langle W_{\ell/L}^q\rangle = (\ell/L)^{\zeta_q} = e^{\log\frac{\ell}{L}\left(qH - \frac{\lambda}{2}q(q-2)\right)},
\end{equation}
which corresponds to the moments of a log-normal distribution. The corresponding Gaussian process $\log  W_{\ell/L}$ is defined by its average $\mu = (H+\lambda)\log\frac{\ell}{L}$ and deviation $\sigma = \sqrt{-\lambda \log\frac{\ell}{L}}$, leading to the distribution:
\begin{equation}
G_{\ell/L}(u)  = \frac{1}{{\sqrt{{-2\pi\lambda\log(\frac{\ell}{L})}}}}e^{\frac{({(H+\lambda)\log(\frac{\ell}{L}) - u)^2}}{{2\lambda\log(\frac{\ell}{L})}}}.
\end{equation}

\section{Parameter estimation of synthetic multifractal fields \label{app:SynthesisAnalysis}}
In the following, we estimate the multifractal parameters of several synthetic multifractal fields.

\begin{figure*}[h]
    \centering
    \includegraphics[width = .9\textwidth]{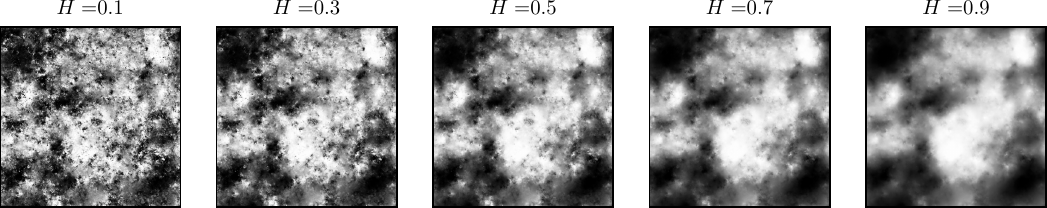}\\
    \vspace{.5cm}
    \includegraphics[width = .8\textwidth]{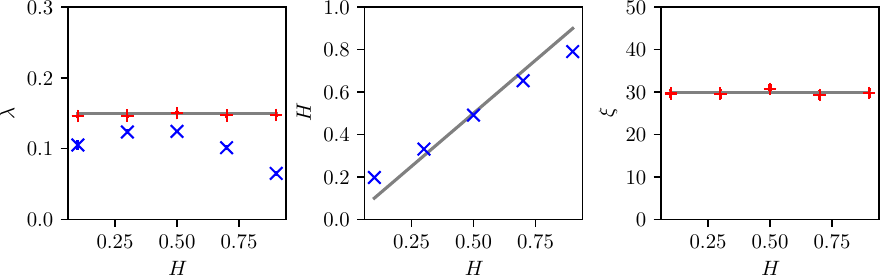}
    \caption{Influence of the roughness exponent $H$. (Top) Quantile representation of simulated fields of size $L = 512$ for $H=\{0.1,0.3,0.5,0.7,0.9\}$, $\lambda = 0.15$ and $\xi_\omega = 30$ pixels. (Bottom) Estimated parameters. Red ($+$) correspond to $C_\omega$ estimations. Blue ($\times$) correspond to $\zeta_q$ estimations. }
    \label{fig:appHurstAnalysis}
    \vspace{1cm}
    \includegraphics[width = .9\textwidth]{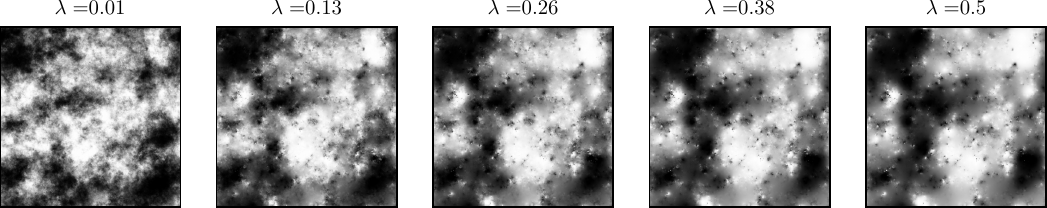}\\
    \vspace{.5cm}

    \includegraphics[width = .8\textwidth]{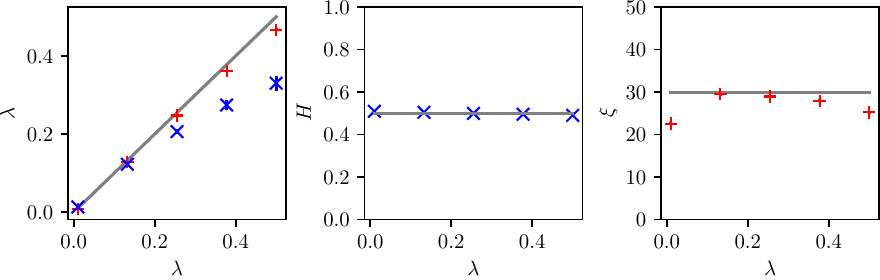}
    \caption{Influence of the intermittency coefficient $\lambda$. (Top) Quantile representation of simulated fields of size $L = 512$ for $\lambda=\{0.01,0.13,0.25,0.38,0.5\}$, $H = 0.5$ and $\xi_\omega = 30$ pixels. (Bottom) Estimated parameters. Red ($+$) correspond to $C_\omega$ estimations. Blue ($\times$) correspond to $\zeta_q$ estimations. }
    \label{fig:appLambdaAnalysis}
\end{figure*}

In Fig.~\ref{fig:appHurstAnalysis}, we modify the Hurst roughness exponent $H$ from $0.1$ to $0.9$. For $H\approx 0.5$, estimations from the scaling exponent spectrum $\zeta_q$ are generally good. For extreme values of $H$ however, the estimated values of $H$ and $\lambda$ deviate from entry values. Note however that this problem can be solved through the power spectrum estimation of the roughness $H = \zeta_2/2$, and the extended self-similarity estimation of $\lambda$~\cite{benzi1993extended,chibbaro_elastic_2016,boudaoud2008observation}. The estimations from $\hat \omega$ are systematically in good agreement.
In Fig.~\ref{fig:appLambdaAnalysis}, we modify $\lambda$ from $0.01$ to $0.50$. For $0<\lambda<.2$, estimations from the scaling exponent spectrum $\zeta_q$ or from the local log-volatility $\hat \omega$ recover good agreement. Note that a low value of $\lambda$ prevents a correct estimation of $\hat \omega$, leading to the discrepancy observed at $\lambda = 0.01$. As $\lambda$ gets higher, the volatility based analysis should be privileged. Note however that $\lambda$ rarely goes beyond $0.25$ in experimental data.

\end{document}